\documentclass[conference,10pt]{IEEEtran}
\usepackage{multirow}
\usepackage{bm}
\usepackage{amsmath}

\usepackage{amsmath, cite}
\usepackage{algpseudocode}
\usepackage{graphicx}
\usepackage{subcaption}

\usepackage{enumitem}

\usepackage{array}
\usepackage{amssymb}
\usepackage{textcomp}
\usepackage{color}
\usepackage{epstopdf}
\usepackage[justification=centering]{caption}
\usepackage{bbm}

\makeatletter

\newcommand{\Rmnum}[1]{\expandafter\@slowromancap\romannumeral #1@}
\renewcommand{\vec}[1]{\ensuremath{\boldsymbol{#1}}}
\makeatother

\begin{document}

\title{An Energy Efficiency Perspective on Massive MIMO Quantization}


\author{\authorblockN{Muris Sarajli\'{c}, Liang Liu, and Ove Edfors}
\authorblockA{Department of Electrical and Information Technology, Lund University, Sweden \\
Email: (muris.sarajlic, liang.liu, ove.edfors)@eit.lth.se}}

\maketitle
\thispagestyle{plain}
\pagestyle{plain}

\begin{abstract}	
One of the basic aspects of Massive MIMO (MaMi) that is in the focus of current investigations is its potential of using low-cost and energy-efficient hardware.
It is often claimed that MaMi will allow for using analog-to-digital converters (ADCs) with very low resolutions and that this will result in overall improvement of energy efficiency.
In this contribution, we perform a parametric energy efficiency analysis of MaMi uplink for the entire base station receiver system with varying ADC resolutions.
The analysis shows that, for a wide variety of system parameters, ADCs with intermediate bit resolutions (4 - 10 bits) are optimal in energy efficiency sense, and that using very low bit resolutions results in degradation of energy efficiency.
\end{abstract}

\IEEEpeerreviewmaketitle


\section{Introduction}

Massive MIMO \cite{Marzetta2010} is an emerging wireless communication technique that promises substantial gains in both spectral and energy efficiency compared to traditional cellular systems.
One feature of MaMi is its robustness to hardware imperfections \cite{Bjornson2015}, which implies that using MaMi will result in improved cost and energy efficiency.
This fact has inspired a flurry of research activity focusing on diverse aspects of hardware impairments in MaMi and specific signaling and signal processing design tailored with the effects of the impairments in mind - all with the common goal of further improving the efficiency of the system.

A significant body of this research aims at analyzing and fine-tuning MaMi-based systems using ADCs with low resolution \cite{Risi2014}, \cite{Jacobsson2015}.
The motivation for such investigations is well-established: the power consumption of ADCs grows at least linearly with sampling rate \cite{Murmann2016} and will prove to be a power consumption bottleneck in systems with very large bandwidths.
Therefore, there is a huge interest in reducing the power consumption of ADCs by reducing the bit resolution as much as possible.

However, it is not perfectly clear whether choosing ADCs with extremely low resolutions will be beneficial from energy efficiency point of view, and analyses of this problem appear scarce.
Only some very recent results \cite{Verenzuela2016}, \cite{Mo2016} seem to indicate that very low bit resolutions are not optimal in energy efficiency sense.

The analysis in this contribution aims at providing an insight in how the overall energy efficiency of a MaMi system - specifically, MaMi uplink - behaves when the ADC resolution is changed.
The total power consumption of MaMi base station is parameterized, so that the analysis covers a wide variety of system architectures - from very economical to very power-hungry.
Also, the ADC power consumption model that is used attempts to reflect the functional dependencies that are found in actual ADC designs and as such aims to be close to hardware design reality.


\section{Energy Efficiency Metric}

Energy efficiency of a base station (BS) receiver in the uplink of a MaMi system is defined as
\begin{equation} 
	\eta = \frac {C}{P_{tot}} \quad \left [ {\text{bits}}/{\text{Joule}} \right ],
\label{eq:energy_efficiency}
\end{equation}
with $C ~ [\text{bits}/s]$ being the uplink sumrate and $P_{tot} ~ [W]$ total power consumption of the MaMi BS (ADCs together with all other receiver blocks, analog and digital).

Dependencies of sumrate and power consumption on ADC bit resolution $b$ need to be resolved separately.
To this end, we first turn to finding an appropriate model for the impact of ADCs on system performance.


\section{ADC Performance Modeling} \label{seq:ADC_performance_modeling}

\begin{figure*}[t] 
   \centering
      \begin{tabular}{c c}
         \includegraphics[width=0.5\textwidth]{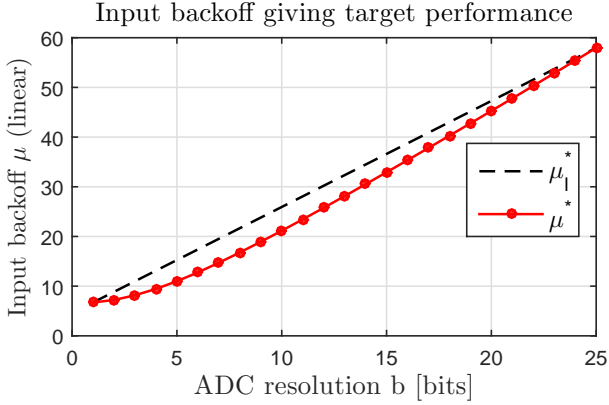} &
         \includegraphics[width=0.5\textwidth]{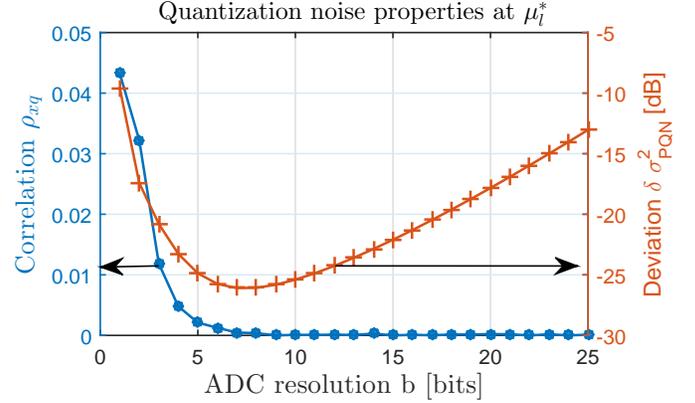}
      \end{tabular}
      \caption{Left: input backoff $\mu$ achieving -13 dB deviation from PQN model. Right: actual deviation and input-distortion correlation for the linear fit}		
			\label{fig:proper_mus_and_performance}
\end{figure*}

This analysis assumes ADCs with bit resolution $b$ that perform scalar quantization and are uniform with $N_q = 2^b$ quantization levels.
Uniform quantization was chosen because it is both close to hardware implementation reality \cite{Gersho1992} and allows for simple and tractable modeling.
Additionally, sampling is assumed to be performed at Nyquist rate.

\emph{Quantizer mapping rule}: given the ADC resolution $b$ and a real positive scalar $X_{ol}$, the \emph{quantization step} of the quantizer is defined as $\Delta = {2 X_{ol}}/{2^b}$, and the values of $N_q$ quantization levels are assigned as $q_i = i \Delta - (N_q + 1) {\Delta}/{2}, ~ i = 1,~\dots~N_q$. 
Additionally, the real line segment $[-X_{ol}, X_{ol}]$ can be divided into $N_q$ equal sub-segments $[-X_{ol}, T_1], [T_1, T_2],~\dots~[T_{N_q-1}, X_{ol}]$ with sub-segment boundaries (thresholds) $T_i = q_i + \frac {\Delta}{2}$.
Given a discrete-time input $x[n]$, the input-output characteristic of the quantizer is defined as
\begin{equation}
	 \text{Q}(x[n]) = 
  \begin{cases} 
   q_1, & \quad x \leq T_1\\
   q_i, & \quad T_{i-1} < x \leq T_i, \quad i = 2, \dots N_q - 1 \\
	 q_{N_q}, & \quad x > T_{N_q - 1}
  \end{cases}
\end{equation}

Quantization $Q(x[n])$ is a nonlinear mapping of $x[n] \in \mathbb{R}$ to a discrete set that results in additive distortion
\begin{equation}
 Q(x[n]) = x[n] + q[n].
\end{equation}
The nature of distortion $q[n]$ is twofold (in the follow-up, time index $n$ is dropped for clarity).
If $|x| > X_{ol}$, $x$ is represented by one of the outer quantization levels $q_1$ or $q_{N_q}$ and we say that the signal is ``clipped''.
Consequently, $q$ is referred to as \emph{clipping} or \emph{overload distortion} with variance $\sigma_{ol}^2$.
On the other hand, if $|x| \leq X_{ol}$, the amplitude of distortion $q$ is bounded by $\frac {\Delta}{2}$; distortion $q$ is then referred to as \emph{granular noise}.

In practical systems, the ADC is usually preceded by an automatic gain control (AGC) variable gain amplifier that is used to conveniently adjust the dynamic range of the signal $x$.
The primary purpose of the AGC is to minimize overload distortion.
A welcome consequence of a properly controlled dynamic range of the input signal is a particularly convenient model for the ADC distortion.

The model in question is described as follows: it was shown in \cite{Sripad1977} that, for a uniform quantizer with normally distributed $x$, the distortion $q$ can very well be approximated as being uniformly distributed, uncorrelated with the input and white, with the variance of $q$ being
\begin{equation}
	\mathbb{E} \{ q^2 \} \approx \frac {1}{3}~X_{ol}^2~2^{-2b} \triangleq \sigma_{\textit{PQN}}^2.
\end{equation}
This commonly used model is usually referred to as the pseudoquantization noise (PQN) model.
The approximations in the PQN model can be made extremely tight if the dynamic range of $x$ is set properly.

A commonly used design parameter for the AGC is the \emph{input backoff} $\mu = {X_{ol}^2}/{\mathbb{E} \{ x^2 \}}$.
In this work, $\mu$ is set so that the normalized deviation of the distortion variance from $\sigma_{\textit{PQN}}^2$, $\delta \sigma_{\textit{PQN}}^2 = | \mathbb{E} \{ q^2 \} - \sigma_{\textit{PQN}}^2|/\sigma_{\textit{PQN}}^2$ is less or equal to -13 dB (indicating that the main source of distortion is granular noise and that it can be safely modeled by the PQN model).
The resulting $\mu^* (b)$ was then approximated linearly by a chord $\mu_l^*(b)$.
Deviation $\delta \sigma_{\textit{PQN}}^2$ and input-distortion crosscorrelation $\rho_{xq} = {\mathbb{E} \{  xq \}}/ \left ( {\sqrt{\mathbb{E} \{ x^2 \}} \sqrt{ \mathbb{E} \{ q^2 \}}} \right )$ were obtained by simulations for $b \in [1, 25]$ and $\mu_l^*(b)$.
The results are shown in Fig. \ref{fig:proper_mus_and_performance} and show that the PQN model applies well even for very low bit resolutions (1 bit) if the AGC backoff is set properly.


\section{System Model and Sumrate Calculation}
By having determined an appropriate model for the effects of ADC, we now employ it in the overall system model for the uplink, which is constructed assuming the following system setup:

\begin{itemize}
	\item Uplink of a single-cell MaMi system with $M$ antennas and $K$ users;
	\item i.i.d. Rayleigh block fading over $T$ symbols;
	\item Least-squares channel estimation is performed using orthogonal pilot sequences of length $\tau$ in the uplink;
	\item Channel estimates are used for linear receiver processing. Maximum ratio combining (MRC) and zero-forcing (ZF) receivers are considered.
\end{itemize}

\begin{figure}[h]
   \centering      
      \includegraphics[width=0.5\textwidth]{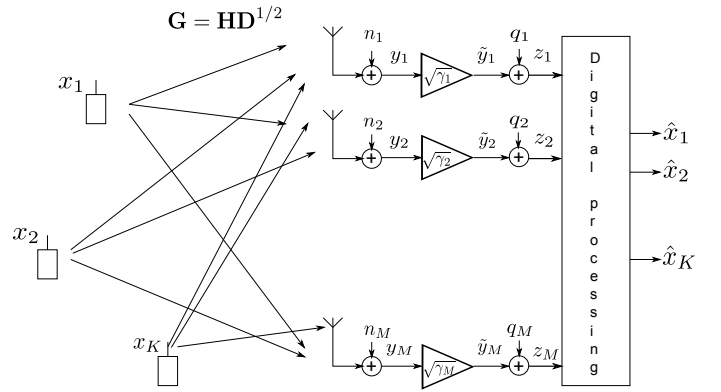}          
   \caption{Uplink system model with quantization noise}
	 \label{fig:system_model}
\end{figure}

A system model of the uplink, where ADCs are substituted by quantization noise sources following the PQN model and AGCs precede ADCs, is illustrated in Fig. \ref{fig:system_model}.

User $k$ sends a data symbol $x_k \in \mathbb{C}$. User symbols are collected in vector $\vec{x} = \left (x_1~x_2~\dots~x_k \right )^T$, with $\mathbb{E} [\vec{x} \vec{x}^H] = \vec{I}_K$.
Single-carrier, narrowband transmission is assumed, and thus the propagation channel is represented by the $M \times K$ matrix $\vec{G} = \vec{HD}^{1/2}$, with the elements of $M \times K$ matrix $\vec{H}$modeling small-scale (SS) fading coefficients.
The elements of $\vec{H}$ are zero mean, circularly symmetric, complex Gaussian random variables with variance 1.

The $K \times K$ matrix $\vec{D}^{1/2}$ is a diagonal matrix of amplitude path gains and large-scale (LS) fading coefficients taken jointly.
The $(m, k)$ element of $\vec{G}$ can be written as $g_{mk} = h_{mk} \sqrt{\beta_k}$, with $h_{mk}$ being the narrowband small-scale fading coefficient between the $k$th user and $m$th antenna and $\beta_k$ the joint path power gain and large-scale fading coefficient.
It should be pointed out that, if some uplink power control is employed, its effects will also be modeled by $\beta$.
In the case of ideal uplink power control, all $\beta_k = 1$.

Assuming that every user transmits with equal transmit power $p_u$, the signal at the receive antennas is
\begin{equation} 
\vec{y} = \sqrt{p_u} \vec{Gx} + \vec{n} = \sqrt{p_u} \vec{HD}^{1/2}\vec{x} + \vec{n}
\label{eq:system_model_at_antennas}
\end{equation}
where $\vec{n} = \left ( n_1~n_2~\dots~n_m \right )^T$ is the vector of input-referred thermal noise at each antenna.
Thermal noise powers at all antennas are assumed equal to $p_n$.

Received signal $y_i$ will experience variations of average power due to SS and LS fading.
This power is averaged over both SSF and LSF and combined with $\mu_l^*$ to find the AGC gains that result in target performance as described in Section \ref{seq:ADC_performance_modeling}. The AGC gain per I/Q branch of $i$th receiver chain is found to be
\begin{equation}
	\gamma_i = \frac {2}{\mu^* \left ( p_u \sum_{k=1}^K \beta_k + p_n \right )}.
\end{equation}
Amplitude AGC gains $\sqrt{\gamma_i}$ can be conveniently collected in a diagonal matrix $\vec{\Gamma}^{1/2}$.

The signal after the AGC is
\begin{align}
	\tilde{\vec{y}} &= \vec{\Gamma}^{1/2} \vec{y}
	= \sqrt{p_u}~\vec{\Gamma}^{1/2} \vec{HD}^{1/2} \vec{x} + \vec{\Gamma}^{1/2} \vec{n} \\ \nonumber
	&= \sqrt{p_u}~\widetilde{\vec{H}} \vec{x} + \tilde{\vec{n}}.
\end{align}

Finally, quantization noise is added.
Assuming $X_{ol} = 1$, variance of complex quantization noise in the $i$th chain is
\begin{equation}
	p_{q, i} = \mathbb{E} \left [ |q_i|^2  \right ] = \frac {2}{3}~2^{-2b_i}.
\end{equation}
and the signal model after the ADC becomes
\begin{equation}
	\vec{z} = \tilde{\vec{y}} + \vec{q} = \sqrt{p_u}~\widetilde{\vec{H}} \vec{x} + \tilde{\vec{n}} + \vec{q},
\end{equation}
where $\vec{q}$ holds the complex quantization noise samples from all antennas. 

\emph{Channel estimation} in the uplink is performed using pilot sequences that are spatially orthogonal and $\tau$ symbols long.
More precisely, pilot sequences for all $K$ users are represented by a $K \times \tau$ matrix $\vec{\Phi} = \sqrt{p_u \tau} \vec{\Psi}$, where in turn $\vec{\Psi}$ is a $K \times \tau$ matrix with orthonormal rows: $\vec{\Psi} \vec{\Psi}^H = \vec{I}_{K \times K}$.
Sequences $\vec{\Phi}$ are optimal for least-squares pilot-based channel estimation \cite{Biguesh2006}. 

When a block of pilot symbols $\vec{\Phi}$ is transmitted, the received signal is
\begin{equation}
	\vec{Z} = \vec{\widetilde{H} \Phi} + \vec{\widetilde{N}} + \vec{\Xi},
\end{equation}
where the columns of matrices $\vec{\widetilde{N}} = \left [ \vec{\tilde{n}_1}~\vec{\tilde{n}_2} \dots \vec{\tilde{n}_\tau} \right ] $ and $\vec{\Xi} = \left [ \vec{q_1}~\vec{q_2} \dots \vec{q_\tau} \right ]$ are thermal and quantization noise vectors for each channel use (symbol).
The least-squares channel estimate is then
\begin{equation} 
	\widehat{\widetilde{\vec{H}}} = \vec{Z} \vec{\Phi}^{\dagger} = {\widetilde{\vec{H}}} + \left ( {\widetilde{\vec{N}}} + \vec{\Xi} \right ) \vec{\Phi}^{\dagger} = {\widetilde{\vec{H}}} + {\widehat{\vec{H}}_\epsilon}.
\label{eq:channel_estimate_w_quantization}
\end{equation}

\emph{Linear processing} matrices for the uplink are formed using the channel estimates:

\begin{itemize}
	\item MRC: $\vec{\hat{A}}_{\text{MRC}} = \vec{\widehat{\widetilde{H}}}$,
	\item ZF: $\vec{\hat{A}}_{\text{ZF}} = \vec{\widehat{\widetilde{H}}} \left ( \vec{\widehat{\widetilde{H}}}^H \vec{\widehat{\widetilde{H}}} \right )^{-1}$.
\end{itemize}

The MIMO receiver applies the processing matrix to estimate the vector of symbols sent by the users as
\begin{equation}
	\hat{\vec{x}} = \vec{\hat{A}}^H \vec{z} = \sqrt{p_u}~\vec{\hat{A}}^H \widetilde{\vec{H}} \vec{x} + \vec{\hat{A}}^H \tilde{\vec{n}} + \vec{\hat{A}}^H \vec{q}.
\end{equation}

It can be shown that $\vec{\hat{A}}$ can be split into a sum of two terms, one being the ``true'' processing matrix (based solely on the actual channel $\vec{\widetilde{H}}$) and the other an error term that is a consequence of channel estimation errors, namely
\begin{itemize}
	\item MRC: $\vec{\hat{A}}_{\text{MRC}} = \vec{A}_{\text{MRC}} + \vec{A}_{\text{MRC},\epsilon} = \vec{\widetilde{H}} + \vec{A}_{\text{MRC},\epsilon}$, and
	\item ZF: $\vec{\hat{A}}_{\text{ZF}} = \vec{A}_{\text{ZF}} + \vec{A}_{\text{ZF},\epsilon} = \vec{\widetilde{H}} \left ( \vec{{\widetilde{H}}}^H \vec{{\widetilde{H}}} \right )^{-1} + \vec{A}_{\text{ZF},\epsilon}$.
\end{itemize}
This fact holds for both MRC (follows directly from (\ref{eq:channel_estimate_w_quantization})) and ZF \cite{Wang2007}.

This simple decomposition allows for splitting the estimate of user data symbol $x_k$, pertaining to $k$th user, into a wanted signal term and a noise term
\begin{equation}
	\hat{x}_k = x_k^{(w)} + w_k = \sqrt{p_u} \vec{a}_k^H \vec{\tilde{h}}_k x_k + w_k,
\end{equation}
where $\vec{a}_k$ is the $k$th column of $\vec{A}$ and $\vec{\tilde{h}}_k$ the $k$th column of $\vec{\widetilde{H}}$.
Additive noise term $w_k$ contains interuser interference and effects of thermal and quantization noise during channel estimation and data transmission phases.

One important observation (the proof of which is omitted here) is that the constituent terms of $w_k$ are all \emph{uncorrelated and Gaussian}.
This is a consequence of several factors, namely: quantization noise being uncorrelated with the input to the ADC, noise in channel estimation phase being independent from the one in data transmission phase, and a large number of antennas (so that the central limit theorem applies).

The signal-to-interference-thermal-and-quantization-noise ratio for $k$th user is then calculated as
\begin{equation}
	\textit{SINQR}_k = \frac {\mathbb{E}_{x,n,q} \{ |x_k^{(w)}|^2 \}}{\mathbb{E}_{x,n,q} \{ |w_k|^2 \}}.
\end{equation}

The ergodic sumrate of the system is the sum of achievable rates for each user, averaged over channel realizations:
\begin{equation}
	C = B \frac {T - \tau}{T} \sum_{k=1}^K \mathbb{E}_{\vec{H}} \left \{  \log_2 (1 + \textit{SINQR}_k) \right \},
\end{equation}
with $B$ being the bandwidth of the system.

\section{Power Consumption Model}

In this work, system setup choices and models aim to be as close to hardware reality as possible.
To this end, we focus on a particular type of ADC - the pipeline ADC.
This type of ADC is typically designed for intermediate bit resolutions, medium to high sampling rates $f_s$ and has power consumption that is comparatively superior to other types of ADCs when observed over a wide range of operating resolutions (very low to very high) \cite{Le2005}, \cite{Svensson2015}, \cite{Murmann2010}.

For the power consumption model of the ADC, this work adopts the model described in \cite{Sundstrom2009}.
It represents a theoretical bound on power dissipation of pipeline ADCs that was nevertheless shown to correctly predict the trends observed in actual designs.
As such, it can be of use in a parametric energy efficiency analysis, where the \emph{character of functional dependency} between $b$ and power consumption is of primary interest.

\begin{figure}[t]
   \centering      
      \includegraphics[width=0.5\textwidth]{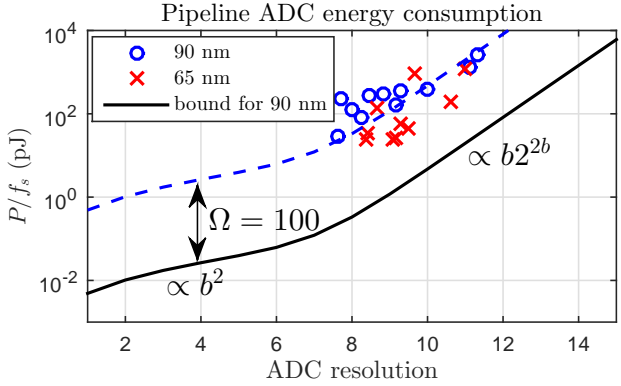}          
   \caption{ADC power consumption model, compared with actual ADC designs}
	 \label{fig:ADC_power_consumption}
\end{figure}

As shown in Fig. \ref{fig:ADC_power_consumption}, where the model from \cite{Sundstrom2009} is compared with selected pipeline ADC designs collected in \cite{Murmann2016}, the functional dependency in the model matches the trend exemplified by state-of-the-art pipeline architectures.
Notwithstanding, there is a gap (about two orders of magnitude wide) between the bound and the designs.
This implies that a correction factor $\Omega$ can be used if the model is to be matched to state-of-the-art; this is illustrated in Fig. \ref{fig:ADC_power_consumption}, where the dashed blue line represents the bound from \cite{Sundstrom2009} offset by $\Omega = 100$.
The power consumption of pipeline ADCs can therefore be calculated as
\begin{equation}
	P_{\textit{ADC}} = \Omega~ \left (c_1 b + c_2 b^2 + c_3 2^{2b} + c_4 b2^{2b}\right ) f_s.
\end{equation}

\begin{figure*}[t] 
   \centering
      \begin{tabular}{c c}
         \includegraphics[width=0.5\textwidth]{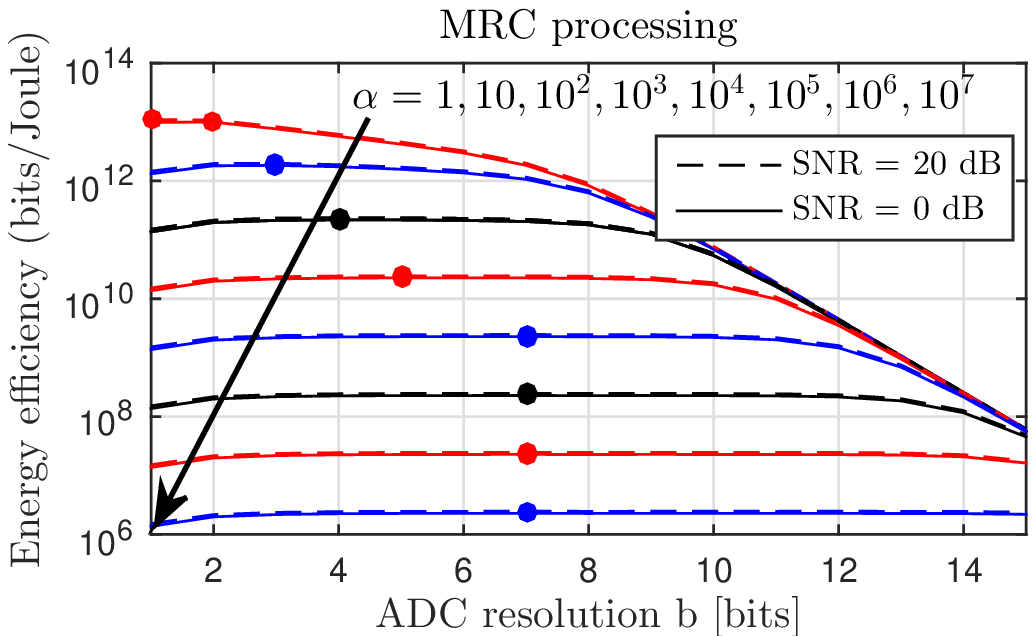} &
         \includegraphics[width=0.5\textwidth]{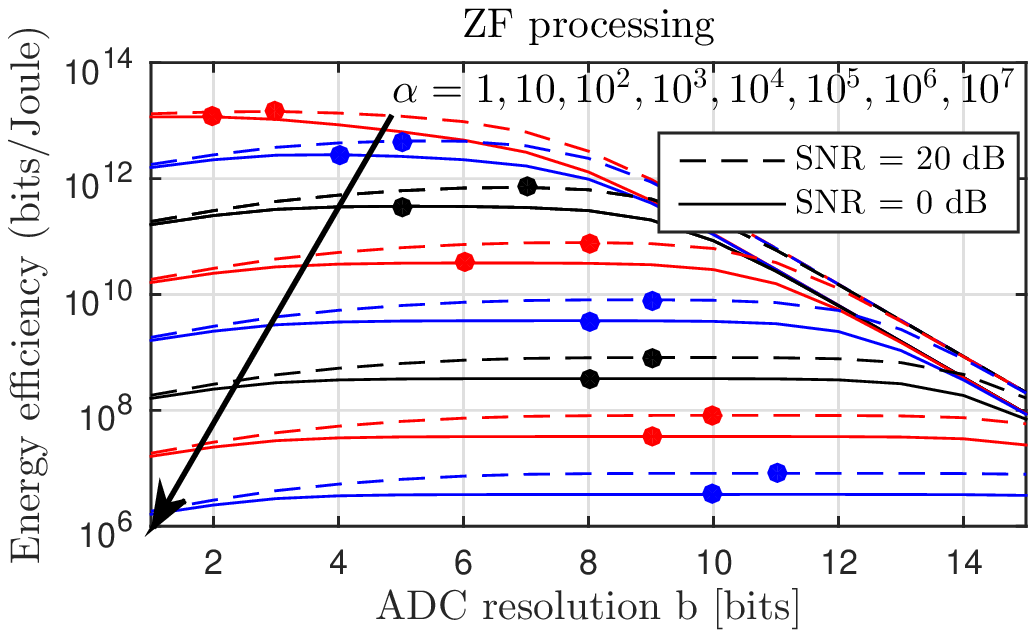}
      \end{tabular}
      \caption{Energy efficiency as a function of architecture parameter $\alpha$ and SNR. Left: MRC, right: ZF}		
			\label{fig:alpha_SNR_sweep}
\end{figure*}

Another characteristic of this model that is worth pointing out is that the power consumption is linear in sampling rate $f_s$.
This trend is also shown to be correct by analyzing actual ADCs in \cite{Murmann2016}; it only breaks down for high sampling rates (on the order of 400 - 500 MHz).
This results in energy efficiency being \emph{independent of system bandwidth}, since passband system bandwidth in the numerator and Nyquist sampling rate of the ADC in the denominator of (\ref{eq:energy_efficiency}) cancel out.

In order to paint the whole picture of energy efficiency of a MaMi base station in the uplink, power consumption of remaining blocks (analog and digital) needs to be taken into account.
This proves to be an extremely challenging task due to wide variability of available system architectures and apparent lack of unifying theoretical information.
Therefore, this work chooses a \emph{parametric approach} to modeling of the total power consumption.

\begin{figure*}
   \centering
      \begin{tabular}{c c}
         \includegraphics[width=0.5\textwidth]{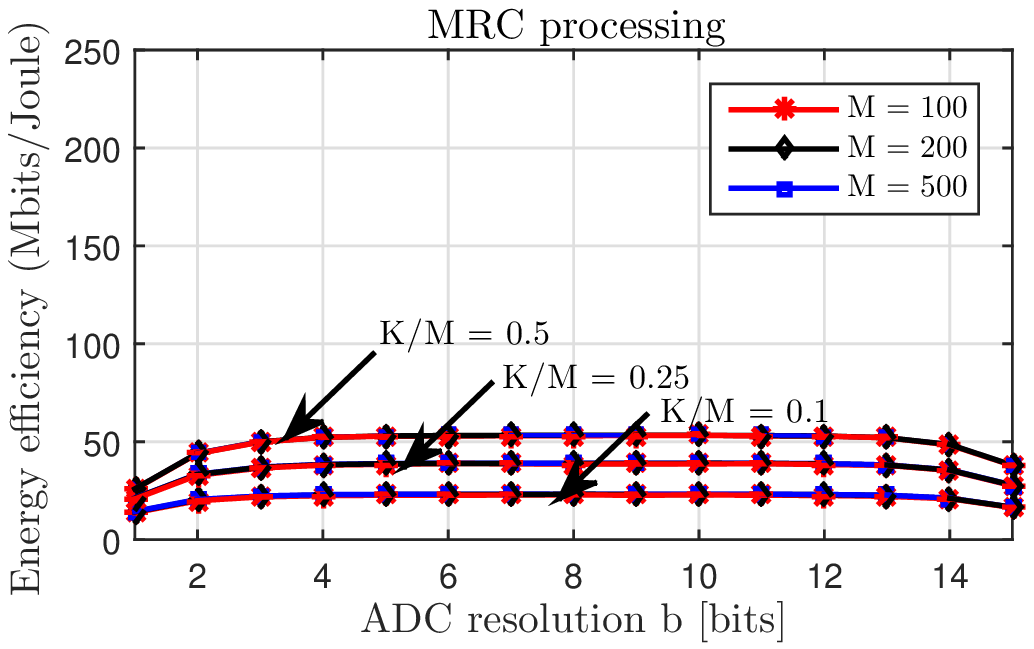} &
         \includegraphics[width=0.5\textwidth]{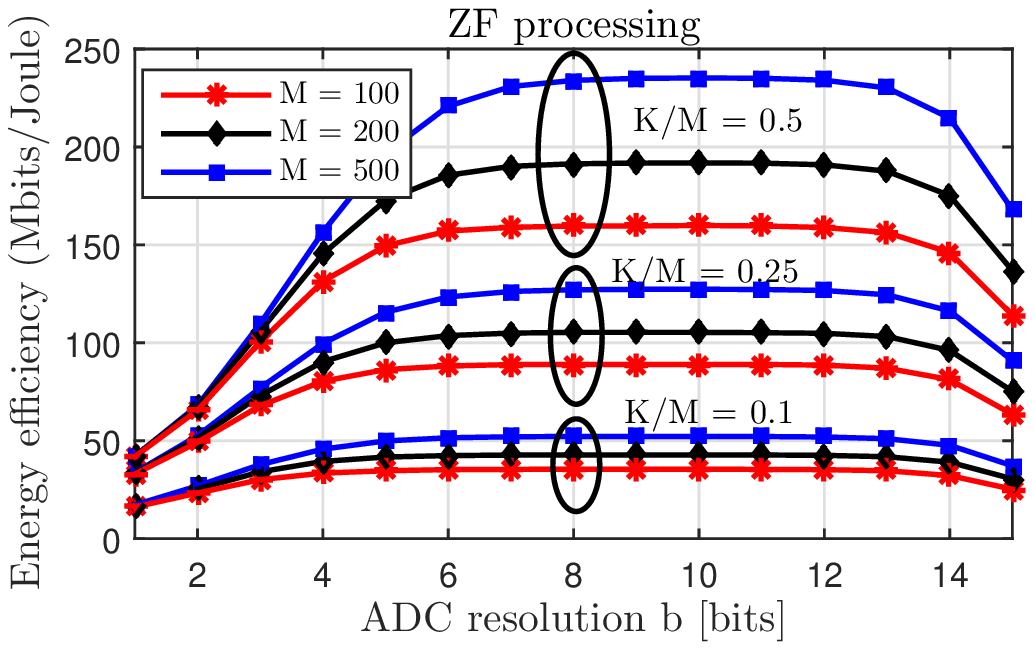}
      \end{tabular}
      \caption{Energy efficiency as a function of $M$, $K$ and $b$. Left: MRC, right: ZF}	
			\label{fig:spatial_loading_M_sweep}
\end{figure*}

\begin{figure*}[t]
   \centering
      \begin{tabular}{c c}
         \includegraphics[width=0.5\textwidth]{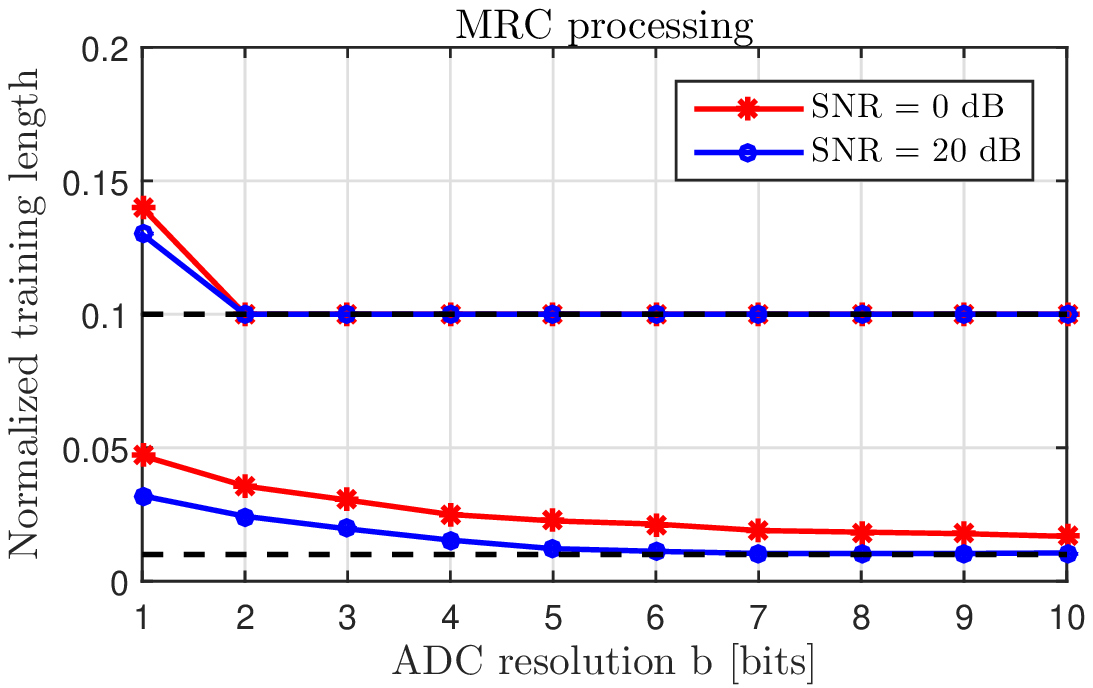} &
         \includegraphics[width=0.5\textwidth]{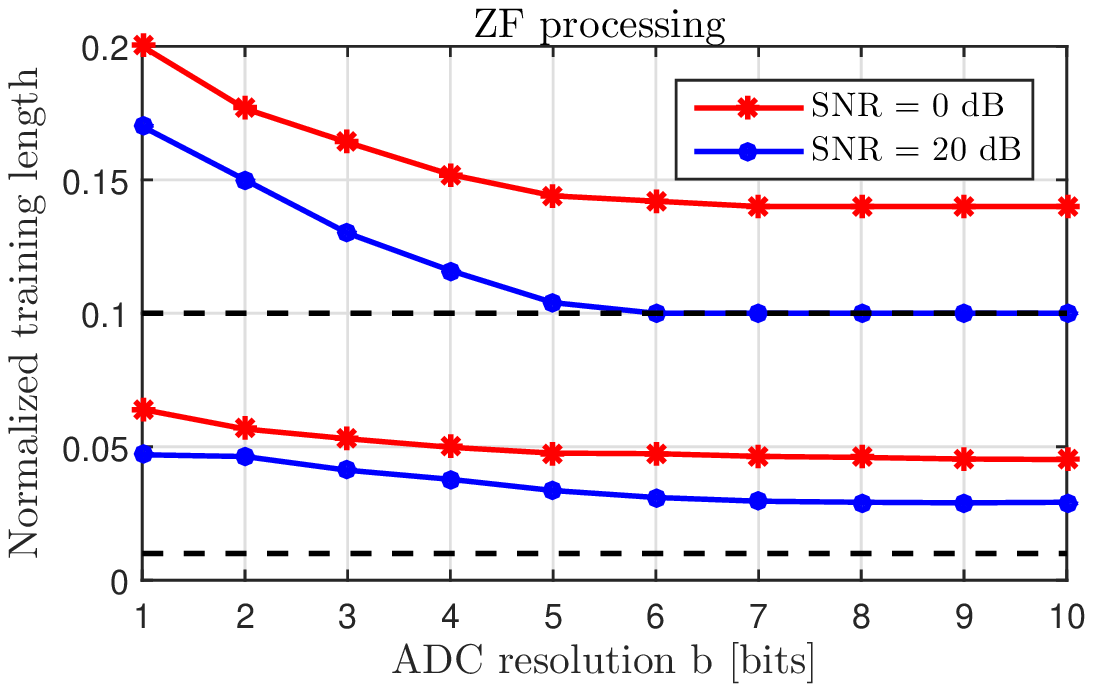}
      \end{tabular}
      \caption{Normalized training length that maximizes energy efficiency. Left: MRC, right: ZF}		
			\label{fig:training_length_sweep}
\end{figure*}

Power consumption of the blocks excluding ADCs, denoted by $P_{\textit{rest}}$, is normalized by $P_{\textit{ADC, ref}}$ - power consumption of ADCs across all RF chains at an arbitrary bit resolution $b_{\textit{ref}}$.
Total power consumption of the BS in the uplink can therefore be expressed as
\begin{equation}
	P_{\textit{tot}} = 2MP_{\textit{ADC}} + P_{\textit{rest}} = 2M(P_{\textit{ADC}} + \alpha P_{\textit{ADC, ref}}),
\end{equation}
where the quantity
\begin{equation}
	\alpha = \frac {P_{\textit{rest}}}{2MP_{\textit{ADC, ref}}}
\end{equation}
is referred to as the \emph{architecture factor}.
Introduction of the architecture factor helps tie together the power consumption of the ADCs and remaining receiver blocks: moreover, it allows for a parameterized analysis that covers a wide range of system architectures.

\section{Results}

The aim of this contribution was to provide an initial overview of the energy efficiency trends as various system parameters change.
To provide this initial insight, system performance simulations have been performed across a wide variety of system parameters.

Alongside primary system parameter $b$, several other important system parameters have been considered, namely $M$, $K$, $T$, $\tau$ and preprocessing $\textit{SNR} = p_u/p_n$ (defined with large-scale fading normalized to the level of thermal noise).
In order to reduce the dimensionality of the analysis, two auxiliary system parameters have been introduced, namely \emph{spatial loading} ($K/M$) and \emph{temporal loading} ($K/T$).

In addition to all the assumptions on system setup listed before, it was assumed that perfect power control was performed in the uplink (so all $\beta_k = 1$).
In all the analyses, reference bit resolution $b_{ref}$ was set to 2.

For the first set of results, $\alpha$ and $\textit{SNR}$ were swept together with $b$.
Additionally, $M = 100$, $\tau = K$, $K/T = 0.01~[\text{users}/\text{coherence time}]$, $K/M = 0.1~[\text{users}/\text{antenna}]$. Results are shown in Fig. \ref{fig:alpha_SNR_sweep}.
Optimal energy efficiency points are denoted by the circular marker.

Results indicate that, as power consumption of ADCs becomes comparable to power consumption of all the other blocks, from energy efficiency point of view it is beneficial to use lower bit resolution. However, in practical system designs it is reasonable to expect that ADC power consumption is only a small fraction of the total power consumption when ADC resolution is low.

Just to provide an illustrative example, BS power model presented in \cite{Desset2014} was used with the parameters listed above (additionally, system bandwidth was assumed to be 20 MHz) and yielded $P_{rest} = 43.3 W$. On the other hand, at $b_{ref} = 2$, using a correction factor $\Omega = 100$, the ADC power consumption model described above gave $2MP_{ADC} = 3~mW$, resulting in $\alpha = 1.5 \times 10^4$. While this is by no means a definite power number, it serves to illustrate what are reasonable orders of magnitude for $\alpha$.

Some other interesting insights can be drawn from this result, for example: system using MRC proves to be quite insensitive to changes in $\textit{SNR}$ and $b$, indicating that an overwhelmingly dominant impairment is the interuser interference and that playing with ADC resolutions will not yield a considerable impact on the energy efficiency; if ZF is used, the dynamics are much more pronounced and show that by going from a system design with a large $\textit{SNR}$ and large $\alpha$ (``wasteful'' system) to a system where $\textit{SNR}$ and $\alpha$ are low (a more ``economical'' system) allows for choosing ADCs with smaller resolutions. Nevertheless, all systems with a ``reasonable'' $\alpha$ (say 10 - $10^5$) should use ADCs with resolutions in the range 4 - 10 bits.

In order to focus more on what are the improvements and degradations of energy efficiency when using different ADC resolutions, we turn to a different analysis where spatial load $K/M$ and $M$ are swept together with $b$, and additionally $\textit{SNR} = 0~dB$, $K/T = 0.01~[\text{users}/\text{coherence time}]$ with $\tau = K$ and $\alpha = 10^4$, results shown in Fig. \ref{fig:spatial_loading_M_sweep}.

What these results show is that going from optimal ADC resolution to a very low one can incur a substantial degradation of the energy efficiency (for ZF processing and with assumed values of system parameters, up to 5.5 times).
This is due to sumrate being degraded while the overwhelming power consumption of other blocks ``drowns'' the coincident savings in power consumption of the ADCs.
Another interesting observation is that, in the ZF case, increasing the number of antennas can help recover the energy efficiency lost by going to lower bit resolutions.

Finally, we take a look at the interplay between the channel estimation length and $b$ in the context of energy efficiency.
We analyze a system with $K/M = 0.1~[\text{users}/\text{antenna}]$, while varying $b$, $\textit{SNR}$ and training length.
Architecture parameter $\alpha$ is again fixed to $10^4$.
What is plotted is the normalized training length $\tau/T$ that maximizes the energy efficiency, results shown in Fig. \ref{fig:training_length_sweep}.
Main takeouts from here are that ZF is much more sensitive to quantization noise during training; even in the case of high temporal loading (indicating fast fading), when there is little room to spare for channel estimation, it is beneficial to train the system longer than minimum required time in order to compensate for the effects of quantization.
The effect becomes more pronounced as the fading becomes slower and channel estimation is not so costly in terms of time.
On the other hand, we see that the system using MRC is so overwhelmed by interuser interference that additional training does little to improve the energy efficiency.

\section{Conclusion}

A parameterized analysis of energy efficiency in the uplink of a MaMi system with varying ADC bit resolutions at the base station has been performed.
System setup and models have been chosen with the aim of being close to practical system implementations.
Initial results (obtained by simulations) indicate that using ADCs with very low bit resolutions is not an optimal approach from energy effiency point of view, except for highly specific system architectures.
Instead, for a wide variety of systems, ADCs with intermediate bit resolutions (4 - 10 bits) are shown to maximize system energy efficiency.
Additionally, it was also shown that systems using MRC uplink processing are quite insensitive to the changes in ADC bit resolution, due to interuser interference being the prime source of impairments in such systems.
On the other hand, systems using ZF processing (in addition to showing overall superiority in terms of energy efficiency compared to MRC - based systems) are shown to be rather sensitive to changes in bit resolution.


\end{document}